# Can the Elephants Handle the NoSQL Onslaught?

Avrilia Floratou
University of Wisconsin-Madison
floratou@cs.wisc.edu

Nikhil Teletia
Microsoft Jim Gray Systems Lab
nikht@microsoft.com

David J. DeWitt
Microsoft Jim Gray Systems Lab
dewitt@microsoft.com

Jignesh M. Patel
University of Wisconsin-Madison
jignesh@cs.wisc.edu

Donghui Zhang[†]
Paradigm4
dzhang@paradigm4.com

## ABSTRACT

In this new era of "big data", traditional DBMSs are under attack from two sides. At one end of the spectrum, the use of document store NoSQL systems (e.g. MongoDB) threatens to move modern Web 2.0 applications away from traditional RDBMSs. At the other end of the spectrum, big data DSS analytics that used to be the domain of parallel RDBMSs is now under attack by another class of NoSQL data analytics systems, such as Hive on Hadoop. So, are the traditional RDBMSs, aka "big elephants", doomed as they are challenged from both ends of this "big data" spectrum? In this paper, we compare one representative NoSQL system from each end of this spectrum with SQL Server, and analyze the performance and scalability aspects of each of these approaches (NoSQL vs. SQL) on two workloads (decision support analysis and interactive data-serving) that represent the two ends of the application spectrum. We present insights from this evaluation and speculate on potential trends for the future.

## 1. INTRODUCTION

The database community is currently at an unprecedented and exciting inflection point. On one hand, the need for data processing products has never been higher than the current level, and on the other hand, the number of new data management solutions that are available has exploded over the past decade. For over four decades, data management typically meant relational data processing, and relational database management systems (RDBMSs) became commonplace in any serious data processing environment. Over the past decade, as industry in nearly every sector of the economy has moved to a data-driven world, there has been an explosion in the volume of data, and the need for richer and more flexible data processing tools.

RDBMSs are no longer the only viable alternative for data-driven applications. First, consider applications in interactive data-serving environments, where consumer-facing artifacts must be computed on-the-fly from a database. Examples of applications in this class include social networks where a consumer-facing web page must be assembled on-the-fly, or a multi-player game in which the objects to be displayed in the next screen must be assembled on-the-fly. Just a few years ago, the standard way to run such applications was to use an RDBMS for the data management component. Now, in such environments, newer NoSQL document systems, such as MongoDB [7], CouchDB [1], Riak [11], etc., are popular alternatives to using an RDBMS. These new NoSQL systems are often designed to have a simpler key-value based data model (in contrast to the relational data model), and are designed to work seamlessly in cluster environments. Thus, many of these systems have in-built "sharding" or partitioning primitives that split large data sets across multiple nodes and keep the shards balanced as new records and/or nodes are added to the system.

This new interactive data-serving domain is largely characterized by queries that read or update a very small amount of the entire dataset. In some sense, one can think of this class of applications as the "new OLTP" domain, bearing resemblance to the traditional OLTP world in which the workload largely consists of short "bullet" queries.

At the other end of the big data application spectrum are analytical decision support workloads that are characterized by complex queries on massive amounts of data. The need for these analytical data processing systems has also been growing rapidly. Once again, (parallel) RDBMSs were largely the only solution for these applications just a few years ago, but now they face competition from another new class of NoSQL systems – namely, systems based on the MapReduce paradigm such as Hive on Hadoop. These NoSQL systems are tailored for large-scale analytics, and are designed specifically to run on clusters of commodity hardware. They assume that hardware/software failures are common, and incorporate mechanisms to deal with such failures. These systems typically also scale easily when adding or removing nodes to an operational cluster.

The question that we ask in this paper is: *How does the performance of RDBMs solutions compare to the NoSQL systems for the two classes of workloads described above, namely interactive data-serving environments and decision support systems (DSS)?* While [19] examined some aspect of this question, it focused only on a small number of simple DSS queries (selected join and aggregate queries). Furthermore, it only considered MapReduce (MR) as an alternative to RDBMSs, and did not consider a more sophisticated MR query processing system like Hive [20]. Furthermore, it has been several years since that effort, and the NoSQL systems have evolved significantly since that time. So, it is interesting to ask how the performance of the NoSQL systems compares to that of parallel RDBMSs today.

---

[†] All of the work by this author on this paper was done while he was at the Microsoft Jim Gray Systems Lab.

In this paper, we present results comparing SQL Server and MongoDB using the YCSB benchmark [14] to characterize how these two SQL and NoSQL systems compare on interactive data-serving environments. We also present results comparing Hive and a parallel version of SQL Server, called PDW, using the TPC-H DSS benchmark [13]. Our results show that the SQL systems currently still have significant performance advantages over both classes of NoSQL systems, but these NoSQL systems are fairly competitive in many cases. The SQL systems will need to continue to keep up their performance advantages and potentially also need to expand their functionality (for example, supporting automatic sharding and a more flexible data model such as JSON) to continue to be competitive.

On a cautionary note, we acknowledge that the evaluation in this paper only considers one data point/system in each class that is considered in this paper, namely a) NoSQL interactive data-serving systems (we use MongoDB), b) MapReduce-based DSS systems (Hive), and c) RDBMS systems (SQL Server and SQL Server PDW). We understand that using other systems in each of these classes may produce different comparative results, and we hope future studies will expand this work to include other systems. In this paper, we have (arguably) taken one representative and leading system in each class, and benchmarked these systems against each other to gather an initial understanding of the emerging big data landscape.

Finally, we note that while we have used some common benchmarks in this paper, the results presented are not audited or official results, and, in fact, were not run in a way that meets all of the benchmark requirements. The results are shown for the sole purpose of providing relative comparisons for this paper, and should not be compared to official benchmark results.

## 2. BACKGROUND
In this section, we describe some background about the different data processing systems that we examine in this paper.

### 2.1 Parallel Data Warehouse (PDW)
SQL Server PDW [6] is a classic shared-nothing parallel database system from Microsoft that is built on top of SQL Server. PDW consists of multiple compute nodes, a single control node and other administrative service nodes. Each compute node is a separate server running SQL Server. The data is horizontally partitioned across the compute nodes. The control node is responsible for handling the user query and generating an optimized plan of parallel operations. The control node distributes the parallel operations to the compute nodes where the actual data resides. A special module running on each compute node called the Data Movement Service (DMS) is responsible for shuffling data between compute nodes as necessary to execute relational operations in parallel. When the compute nodes are finished, the control node handles post-processing and re-integration of results sets for delivery back to the users.

### 2.2 Hive
Hive [3] is an open-source data warehouse built on top of Hadoop [2]. It provides a structured data model for data that is stored in the Hadoop Distributed Filesystem (HDFS), and a SQL-like declarative query language called HiveQL. Hive converts HiveQL queries to a directed acyclic graph of MapReduce jobs, and thus saves the user from having to write the more complex MapReduce jobs directly.

Data organization in Hive is similar to that found in relational databases. Starting from a coarser granularity, data is stored in *databases, tables, partitions* and *buckets*. More details about the data layout in Hive are provided in Section 3.3.2.

Finally, Hive has support for multiple data storage formats including text files, sequence files, and RCFiles [17]. Users can also create custom storage formats as well as serializers/deserializers, and plug them into the system.

### 2.3 MongoDB
MongoDB [7] is a popular open-source NoSQL database. Some of its features are a document-oriented storage layer, indexing in the form of B-trees, auto-sharding and asynchronous replication of data between servers.

In MongoDB data is stored in *collections* and each collection contains *documents*. Collections and documents are loosely analogous to tables and records, respectively, found in relational databases. Each document is serialized using BSON. MongoDB does not require a rigid schema for the documents. Specifically, documents in the same collection can have different structures.

Another important feature of MongoDB is its support for auto-sharding. With sharding, data is partitioned amongst multiple nodes in an order-preserving manner. Sharding is similar to the horizontal partitioning technique that is used in parallel database systems. This feature enables horizontal scaling across multiple nodes. When some nodes contain a disproportionate amount of data compared to the other nodes in the cluster, MongoDB redistributes the data automatically so that the load is equally distributed across the nodes/shards.

Finally, MongoDB supports failover via *replica sets*, which is its mechanism for implementing asynchronous master/slave replication. A replica set consists of two or more nodes that are copies of each other. More information about the semantics of replica sets can be found in [8]. In the following sections, we use the name **Mongo-AS** (MongoDB with auto-sharding) when referring to the original MongoDB implementation.

### 2.4 Client-side Sharded SQL Server and MongoDB
For our experiments, we created a SQL Server implementation (**SQL-CS**) that uses client-side hashing to determine the home node/shard for each record by modifying the client-side application that runs the YCSB benchmark. We implemented this client-side sharding so that we could compare MongoDB(-AS) with SQL Server in a *cluster* environment. We also took the client-side sharding code and implemented it on top of MongoDB. This implementation of client-side sharding on MongoDB is denoted as **MongoDB-CS**, allowing us to compare MongoDB-AS with MongoDB-CS (and SQL-CS).

We note that both SQL-CS and Mongo-CS do not support some of the features that are supported by Mongo-AS. First, whereas Mongo-AS uses a form of range partitioning to distribute the records across the shards, the Mongo-CS and SQL-CS implementations both use hash partitioning. Another difference is that the Mongo-CS implementation does not use any of the routing (mongos), configuration (config db), and balancer processes that are part of Mongo-AS. As a result, load balancing cannot happen automatically as in Mongo-AS, where the auto-sharding mechanism aims to continually balance the load across all the nodes in the cluster. However, Mongo-CS makes use of the basic "mongod" process, which is responsible for processing the client's requests. Finally, Mongo-CS and SQL-CS do not support automatic failover. We note that these features listed above were not the key subject of performance testing in the benchmark (YCSB) that we use in this paper.

On the flip side, we also note that SQL Server has many features that are not supported in MongoDB. For example, MongoDB has a flexible data model that makes it far easier to deal with schema changes. MongoDB also supports read/write atomic operations on single data entities, whereas SQL Server provides full ACID semantics and multiple isolation levels. SQL Server also has better manageability and performance analysis tools (e.g. database tuning advisor).

## 3. EVALUATION
In this section, we present an experimental evaluation of a RDBMS and a NoSQL system on a DSS and a "modern" OLTP workload. More specifically, we use TPC-H [13] to evaluate Microsoft's Parallel Data Warehouse and Hive. We also compare MongoDB (Mongo-AS) with both client-side sharded Microsoft SQL Server (SQL-CS) and MongoDB (Mongo-CS) implementations, using the YCSB benchmark. The following sections present details about the hardware and the software configuration that is used in our experiments.

## 3.1 Hardware Configuration
All experiments were run on a cluster of 16 nodes connected by 1Gbit HP Procurve 2510G 48 (J9280A) Ethernet switch. Each node has dual Intel Xeon L5630 quad-core processors running at 2.13 GHz, 32 GB of main memory, and 10 SAS 10K RPM 300GB hard drives. One of the hard drives is always reserved for the operating system.

When evaluating PDW and Hive, we used eight disks to store the data. These disks were organized as one RAID 0 volume when the system was running Hive, and configured as separate logical volumes when running PDW. The log data for each PDW node was stored on a separate hard disk.

For Hive, we used one extra node to run the namenode and the jobtracker processes only. PDW needs two extra nodes, used as a control node and as a landing node respectively. The landing node is responsible for data loading and does not participate in query execution. All the extra nodes were connected to the same Ethernet switch that is used by the remaining 16 nodes in the cluster. The operating system was Windows Server 2008 R2 when running PDW, and Ubuntu 11.04 when running Hive.

For the YCSB benchmark experiments, eight nodes were used as servers (running SQL or Mongo) and eight were used to run the client benchmark. Similar to the DSS experimental setting, eight disks were used to store the data for the OLTP experiments. These disks were configured as RAID 0 when running MongoDB, and were treated as separate logical volumes when running SQL Server. The log data for SQL Server was stored on a separate hard disk. For the Mongo-AS experiments, we used one extra node as the "config" server. The "config" server keeps metadata about the cluster's state. The operating system was, in both cases, Windows Server 2008 R2.

## 3.2 Software Configuration
In this section, we describe the software configuration for each system that we tested.

### 3.2.1 Hive and Hadoop
We used Hive version 0.7.1 running on Hadoop version 0.20.203. We configured Hadoop to run 8 map tasks and 8 reduce tasks per node (a total of 128 map slots and 128 reduce slots). The maximum JVM heap size was set to 2GB per task. We used a 256 MB HDFS block size, and the HDFS replication factor was set to 3. The TPC-H Hive scripts are available online [12]. However, since Hive now supports features that were not available when these scripts were written, we modified the scripts in the following ways:

1. Instead of using text files to store the data, we used the RCFile format [17]. The RCFile layout is considered to be faster than a row-store format (e.g. text file, sequence file) since it can eliminate some I/O operations [16], [17]. All the TPC-H base tables are stored in compressed (GZIP) RCFile format. Some TPC-H queries were split manually (by the Hive team) into smaller sub-queries, since HiveQL is not expressive enough to support the full SQL-92 specification; the output of these intermediate queries is also stored in the RCFile format.
2. We enabled the map-side aggregation, map-side join and bucketed map-join features of Hive, which usually improves performance by avoiding executing the reduce phase of a MapReduce job.
3. We set the number of reducers for each MapReduce job to the total number of reduce slots in the cluster (128 reducers). We found that this setting significantly improves the performance of Hive when running the TPC-H benchmark, since all the reducers can now complete in one reduce round.

Finally, all the results produced by the map tasks are compressed using LZO, according to the suggestions of the Hive team [12] for appropriate setting of Hive when running TPC-H.

### 3.2.2 PDW
For our experiments we used a pre-release version (November 2011) of PDW AU3. Each compute node runs SQL Server 2008 configured to use a maximum of 24GB of memory for its buffer pool. Each compute node contains 8 horizontal data partitions (a total of 128 partitions across the cluster).

SQL Server PDW is only sold as an appliance. In an appliance each node is configured much larger amounts of memory and storage and the nodes are interconnected using Infiniband, and not Ethernet. Hence, the results a customer would see would be much faster than what we report below for an appliance with a similar number of nodes. Since, we wanted to avoid an apples-to-oranges comparison between PDW and Hive we used exactly the same hardware for both systems.

### 3.2.3 MongoDB (Mongo-AS)
We used MongoDB version 1.8.2. MongoDB supports auto-sharding so that it can scale horizontally across multiple nodes. In our configuration, the data is spread across 128 shards. We ran 16 "mongod" processes on each one of our 8 server machines. Each "mongod" process is responsible for one shard.

In MongoDB (version 1.8.2) any number of concurrent read operations are allowed, but a write operation can block all other operations. That is because MongoDB uses a global lock for writes (there is one such lock per "mongod" process)[‡]. Consequently, we chose to run 16 processes per server node instead of one. In this way, we can exploit the fact that our nodes have 16 cores (hyper-threaded) and, at the same time, increase the concurrency when the workload contains inserts or updates. Our single node experiments have shown that running 16 processes per machine has better performance than running one or eight processes when using the YCSB benchmark.

**Table 1. Data layout in Hive and PDW**

| Table | HIVE Partition Column | HIVE Buckets | PDW Partition Column | PDW Replication |
|---|---|---|---|---|
| Customer | c_nationkey | 8 buckets per partition on c_custkey | c_custkey | No |
| Lineitem | -- | 512 buckets on l_orderkey | l_orderkey | No |
| Nation | -- | -- | -- | Yes |
| Orders | -- | 512 buckets on o_orderkey | o_orderkey | No |
| Part | -- | 8 buckets on p_partkey | p_partkey | No |
| Partsupp | -- | 8 buckets on ps_partkey | ps_partkey | No |
| Region | -- | -- | -- | Yes |
| Supplier | s_nationkey | 8 buckets per partition on s_suppkey | s_suppkey | No |

Except for the "config db" and "mongod" processes, we launched 8 "mongos" processes, one at each server machine. The "mongos" process is responsible for routing client requests to the appropriate "mongod" instance. All the clients that run on the same client node connect to the same "mongos" process. Since we have 8 client nodes, there is a "1-1" correspondence between the "mongos" processes and the client nodes.

MongoDB supports failover by using a form of asynchronous master/slave replication, called replica sets. For our experiments, we did not create any replica sets.

## 3.3 Traditional DSS Workload: Hive vs. PDW

In this section we describe the DSS TPC-H workload and various parameters related to this workload for both PDW and Hive.

### 3.3.1 Workload Description

We used TPC-H at four scale factors (250 GB, 1000 GB, 4000 GB, 16000 GB) to evaluate the performance of PDW and Hive. These four scale factors represent cases where different portions of the TPC-H tables fit in main memory. We noticed that the TPC-H generator does not produce correct results at the 16000 scale factor (this scale factor cannot be reported in the official benchmark results). More specifically, the values generated for the *partkey* and *custkey* fields in the *mk_order* function are negative numbers. These numbers are produced using the RANDOM function, which overflows at the 16TB scale. Hence, we modified the generator code to use a 64-bit random number generator (RANDOM64). For all the scale factors, we executed the 22 TPC-H queries that are included in the benchmark, sequentially. We didn't execute the two TPC-H refresh functions, because the Hive version that we used, does not support deletes and inserts into existing tables or partitions (the newer Hive versions 0.8.0 and 0.8.1 do support INSERT INTO statements).

### 3.3.2 Data Layout

A Hive table can contain *partitions* and/or *buckets*. In Hive, each partition corresponds to one HDFS directory and contains all the records of the table that have the same value on the partitioning attributes. Selection queries on the partitioning columns can benefit from this layout since only the necessary partitions are scanned instead of the whole table.

A Hive table can also consist of a number of buckets. A bucket is stored in a file within the partition's or table's directory depending on whether the table is a partitioned table or not. The user provides a bucketing column as well as the number of buckets that should be created for the table. Hive determines the bucket number for each row of the table by hashing on the value of the bucketing column. Each bucket may contain rows with different values on the bucketing column. During a join, if the tables involved are bucketed on the join column, and the buckets are a multiple of each other, the buckets can be joined with each other in a map-side join only.

As seen in Table 1, a Hive table can contain both partitions and buckets (e.g. *Customer* table in Table 1). In this case the table consists of a set of directories (one for each partition). Each directory contains a set of files, each one corresponding to one bucket. Hive tables can also be only *partitioned* or only *bucketed* (e.g. *Lineitem* table in Table 1). In the first case, selection queries on the partitioning columns can benefit from the layout. However, join queries cannot benefit unless there is a predicate on the partitioning columns on at least one of the tables. Bucketed tables can help improve the performance of joins but cannot improve selection queries even if there is a selection predicate on the bucketing column.

In PDW, a table can be either horizontally partitioned or replicated across all the nodes. When partitioned, the records are distributed to the partitions using a hash function on a partition column. Table 1 summarizes the data layouts for Hive and PDW.

As can be seen in Table 1, the bucket columns used in Hive are the same as those used to horizontally partition the PDW tables. Each bucket is also sorted on the corresponding bucket column. The PDW tables consist of 128 partitions (8 data "distributions" per node).

Previous work [19] has shown that one of the major reasons why relational databases outperform Hadoop on some workloads is their inherent indexing support. For our experiments we decided not to use any type of index for the PDW tables (including primary key indices). The reason behind this decision is that the Hive version we used does not support automatic generation of query plans that consider the available indices. Instead, the user has to rewrite the query so that it takes into account the appropriate indices. This process quickly becomes complicated with complex queries like those in the TPC-H benchmark since the user has to manually produce the "optimal" query plan. The newer versions of Hive, have improved their index support and there is an ongoing effort on seamlessly integrating indexing in

---

[‡] MongoDB (version 2.0) implements yield-on-page-fault and yield-on-long-operation features, which potentially will allow for more concurrency, but our testing found it unreliable.

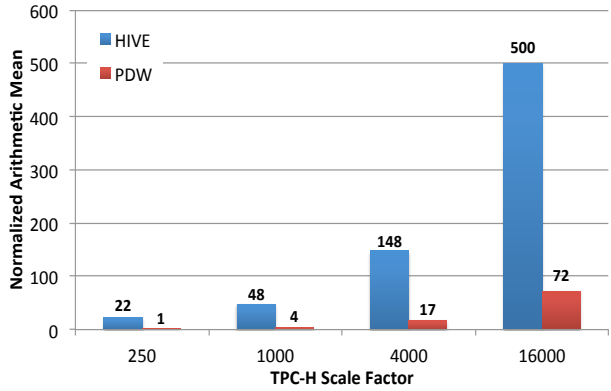 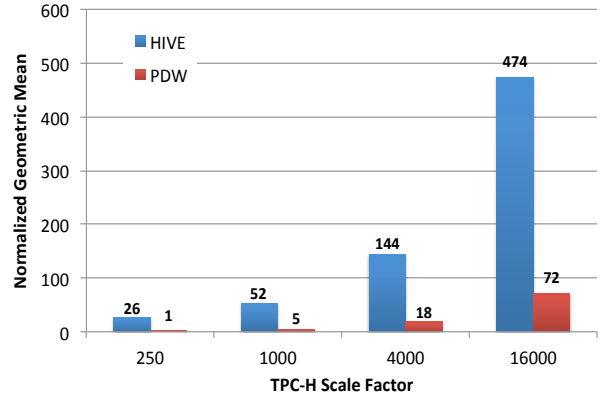

(a) Normalized arithmetic mean          (b) Normalized geometric mean

**Figure 1. TPC-H Performance on HIVE and PDW at different scale factors (normalized to PDW at Scale Factor = 250). See Table 3 for the detailed numbers.**

Hive (e.g. [4],[5]). As part of future work, we plan on comparing the performance of PDW with Hive, once Hive's optimizer starts considering indices.

### 3.3.3 Data Preparation and Load Times

In this section, we describe the data preparation steps for each system. We also present the data loading time for each system.

For Hive, we generated the TPC-H dataset in parallel across the 16 nodes of our cluster using the TPC "dbgen" program. All the data is stored on a separate hard disk that is not used to store HDFS data files. Before starting the loading process, we created one Hive table for each TPC-H table. In the table definition, we provide the schema of the table, the partitioning and bucketing columns (if applicable) and the storage format (RCFile).

Data is loaded into Hive using two phases. First, the TPC-H data files are loaded on each node in parallel, directly into HDFS, as plain text using the HDFS command-line utility that copies data from the local filesystem to HDFS. For each TPC-H table, an external Hive table is then created. The table points to a directory in HDFS that contains all the relevant data for this table. In the next phase, the data is converted from the text format into the compressed RCFile format. This conversion is done using a Hive query that selects all the tuples of each external table and inserts them into the corresponding Hive table.

When loading into PDW, the TPC-H data is generated on the landing node. Before loading the data, the necessary TPC-H tables are created by using the CREATE TABLE statement and specifying the schema and distribution of the tables (replicated or hash-distributed). The generated data is loaded using the "dwloader" utility of PDW, which splits the text files that are generated at the landing node, into multiple chunks. These chunks are then loaded to the 16 compute nodes of the cluster in parallel.

Table 2 presents the data loading times for both systems.

**Table 2. Load times for Hive and PDW**

|      | Load Time (in minutes) | | | |
| --- | --- | --- | --- | --- |
|      | 250 GB | 1 TB | 4 TB | 16 TB |
| **HIVE** | 38 | 125 | 519 | 2512 |
| **PDW**  | 79 | 313 | 1180 | 4712 |

### 3.3.4 Experimental Evaluation

In this section we present an analysis of the performance and scalability aspects of Hive and PDW when running TPC-H.

Table 3 presents the running time of the queries in PDW and Hive for each TPC-H query for each of the four scale factors. For each scale factor, the table contains the speedup of PDW over Hive. The last six columns, display a scaling factor for each TPC-H query. This factor shows the increase in the execution time of the query when the TPC-H scale factor is increased by a factor of four. The table also contains the arithmetic and geometric mean of the response times at all scale factors. The values of *AM-9* and *GM-9* correspond to the arithmetic and geometric mean of all the queries but Query 9, since Query 9 did not complete in Hive at the 16TB scale factor due to lack of disk space.

#### 3.3.4.1 Performance Analysis

Figure 1 presents an overview of the performance results for the TPC-H queries at the four tested scale factors for both Hive and PDW (the detailed numbers are in shown in Table 3). Figure 1 (a) shows the normalized arithmetic mean of the response times for the TPC-H queries, and Figure 1 (b) shows the normalized geometric mean of the response times; the numbers plotted in Figure 1 are normalized to the response times for PDW at scale factor 250. These numbers were computed based on the *AM-9* and *GM-9* values.

As shown in the figure, PDW has a significantly lower (normalized) arithmetic and geometric mean. Moreover, PDW is always faster than Hive for all TPC-H queries and at all scale factors (see Table 3). The average speedup of PDW over Hive is greater for small datasets (34.1X for the 250 GB scale factor). This behavior can be attributed to two factors: a) PDW can better exploit the property that, for small scale factors, most of the data fits in memory, and b) As we will discuss below, Hive has high overheads for small datasets.

In this section, we analyze two TPC-H queries in which PDW significantly outperformed Hive at all scale factors, to gather some insights into the performance differences.

### Query 5

As shown in Table 3, Query 5 is approximately 19 times faster on PDW than Hive on the 16TB scale. This query joins six tables

Table 3. Performance of Hive and PDW on TPC-H at four scale factors

| Query | SF = 250 GB Time(sec) | | | SF = 1000 GB Time(sec) | | | SF = 4000 GB Time(sec) | | | SF = 16000 GB Time (sec) | | | PDW | | | HIVE | | |
|---|---|---|---|---|---|---|---|---|---|---|---|---|---|---|---|---|---|---|
| | HIVE | PDW | Speedup | HIVE | PDW | Speedup | HIVE | PDW | Speedup | HIVE | PDW | Speedup | 250 → 1000 | 1000 → 4000 | 4000 → 16000 | 250 → 1000 | 1000 → 4000 | 4000 → 16000 |
| Q1 | 207 | 54 | 3.8 | 443 | 212 | 2.1 | 1376 | 864 | 1.6 | 5357 | 3607 | 1.5 | 3.9 | 4.1 | 4.2 | 2.1 | 3.1 | 3.9 |
| Q2 | 411 | 7 | 58.7 | 530 | 25 | 21.2 | 1081 | 115 | 9.4 | 3191 | 495 | 6.4 | 3.6 | 4.6 | 4.3 | 1.3 | 2.0 | 3.0 |
| Q3 | 508 | 32 | 15.9 | 1125 | 112 | 10.0 | 3789 | 606 | 6.3 | 11644 | 2572 | 4.5 | 3.5 | 5.4 | 4.2 | 2.2 | 3.4 | 3.1 |
| Q4 | 367 | 8 | 45.9 | 855 | 54 | 15.8 | 2120 | 187 | 11.3 | 6508 | 629 | 10.3 | 6.8 | 3.5 | 3.4 | 2.3 | 2.5 | 3.1 |
| Q5 | 536 | 33 | 16.2 | 1686 | 80 | 21.1 | 5481 | 253 | 21.7 | 19812 | 1060 | 18.7 | 2.4 | 3.2 | 4.2 | 3.1 | 3.3 | 3.6 |
| Q6 | 79 | 5 | 15.7 | 166 | 41 | 4.0 | 537 | 142 | 3.8 | 2131 | 526 | 4.1 | 8.2 | 3.5 | 3.7 | 2.1 | 3.2 | 4.0 |
| Q7 | 1007 | 19 | 53.0 | 2447 | 80 | 30.6 | 7694 | 240 | 32.1 | 24887 | 955 | 26.1 | 4.2 | 3.0 | 4.0 | 2.4 | 3.1 | 3.2 |
| Q8 | 967 | 9 | 107.4 | 2003 | 89 | 22.5 | 6150 | 238 | 25.8 | 18112 | 814 | 22.3 | 9.9 | 2.7 | 3.4 | 2.1 | 3.1 | 2.9 |
| Q9 | 2033 | 207 | 9.8 | 7243 | 844 | 8.6 | 27522 | 3962 | 6.9 | -- | 15494 | -- | 4.1 | 4.7 | 3.9 | 3.6 | 3.8 | -- |
| Q10 | 489 | 14 | 35.0 | 1107 | 67 | 16.5 | 2958 | 265 | 11.2 | 13195 | 981 | 13.5 | 4.8 | 4.0 | 3.7 | 2.3 | 2.7 | 4.5 |
| Q11 | 242 | 3 | 80.8 | 258 | 18 | 14.3 | 695 | 99 | 7.0 | 1964 | 302 | 6.5 | 6.0 | 5.5 | 3.1 | 1.1 | 2.7 | 2.8 |
| Q12 | 253 | 5 | 50.6 | 490 | 44 | 11.1 | 1597 | 192 | 8.3 | 5123 | 631 | 8.1 | 8.8 | 4.4 | 3.3 | 1.9 | 3.3 | 3.2 |
| Q13 | 392 | 51 | 7.7 | 629 | 190 | 3.3 | 1428 | 772 | 1.8 | 4577 | 3061 | 1.5 | 3.7 | 4.1 | 4.0 | 1.6 | 2.3 | 3.2 |
| Q14 | 154 | 7 | 22.0 | 353 | 64 | 5.5 | 769 | 164 | 4.7 | 2556 | 640 | 4.0 | 9.1 | 2.6 | 3.9 | 2.3 | 2.2 | 3.3 |
| Q15 | 444 | 21 | 21.1 | 585 | 99 | 5.9 | 1145 | 377 | 3.0 | 2768 | 1397 | 2.0 | 4.7 | 3.8 | 3.7 | 1.3 | 2.0 | 2.4 |
| Q16 | 460 | 36 | 12.8 | 654 | 71 | 9.2 | 1732 | 223 | 7.8 | 5695 | 549 | 10.4 | 2.0 | 3.1 | 2.5 | 1.4 | 2.6 | 3.3 |
| Q17 | 654 | 93 | 7.0 | 1717 | 406 | 4.2 | 6334 | 1679 | 3.8 | 25662 | 6757 | 3.8 | 4.4 | 4.1 | 4.0 | 2.6 | 3.7 | 4.1 |
| Q18 | 786 | 20 | 39.3 | 2249 | 103 | 21.8 | 8264 | 482 | 17.1 | 25964 | 2880 | 9.0 | 5.2 | 4.7 | 6.0 | 2.9 | 3.7 | 3.1 |
| Q19 | 376 | 16 | 23.5 | 1069 | 73 | 14.6 | 4005 | 272 | 14.7 | 17644 | 958 | 18.4 | 4.6 | 3.7 | 3.5 | 2.8 | 3.7 | 4.4 |
| Q20 | 606 | 20 | 30.3 | 1296 | 101 | 12.8 | 2461 | 425 | 5.8 | 11041 | 1611 | 6.9 | 5.1 | 4.2 | 3.8 | 2.1 | 1.9 | 4.5 |
| Q21 | 1431 | 31 | 46.1 | 3217 | 138 | 23.3 | 13071 | 927 | 14.1 | 40748 | 4736 | 8.6 | 4.5 | 6.7 | 5.1 | 2.2 | 4.1 | 3.1 |
| Q22 | 908 | 19 | 47.8 | 1145 | 71 | 16.1 | 1744 | 255 | 6.8 | 3402 | 1270 | 2.7 | 3.7 | 3.6 | 5.0 | 1.3 | 1.5 | 2.0 |
| AM | 605 | 32 | 34.1 | 1421 | 136 | 13.4 | 4634 | 579 | 10.2 | -- | 2360 | -- | 5.1 | 4.0 | 3.9 | 2.1 | 2.9 | -- |
| GM | 474 | 19 | 25.2 | 971 | 89 | 10.9 | 2727 | 352 | 7.7 | -- | 1368 | -- | 4.7 | 3.9 | 3.9 | 2.0 | 2.8 | -- |
| AM-9 | 537 | 24 | 35.3 | 1144 | 102 | 13.6 | 3544 | 418 | 10.4 | 11999 | 1735 | 9.0 | 5.2 | 4.0 | 3.9 | 2.1 | 2.9 | 3.4 |
| GM-9 | 442 | 17 | 26.3 | 882 | 80 | 11.0 | 2443 | 314 | 7.8 | 8062 | 1219 | 6.6 | 4.8 | 3.9 | 3.9 | 2.0 | 2.8 | 3.3 |

(*customer*, *orders*, *lineitem*, *supplier*, *nation* and *region*) and then performs an aggregation.

The plans produced by the PDW and the Hive query optimizers are as follows:

**PDW**: PDW first shuffles the *orders* table on o_*custkey*. The shuffle is completed after approximately 258 seconds. Then, PDW performs a join between the *customer*, *orders*, *nation* and *region* tables. This join can be performed locally on each PDW node since the *nation* and *region* tables are replicated across all the nodes of the cluster and the *customer* table is hash partitioned on the *c_custkey* attribute. The output of the join is shuffled on the *o_orderkey* attribute. The join and shuffle phases run for approximately 86 seconds. The table produced by the previous operations is locally joined with the *lineitem* table, which is partitioned on the *l_orderkey* attribute and then shuffled on the *l_suppkey* attribute. This shuffle and join phase runs for 665 seconds. Then, the resulting table is joined with the *supplier* table (locally). During this join, a partial aggregation on the *n_name* attribute is performed. Finally, all the local tables produced at each PDW node are globally aggregated on the *n_name* attribute to produce the final result. The join, the partial aggregation and the global aggregation operations complete after 40 seconds.

**Hive**: Hive first performs a map-side join between the *nation* and the *region* tables. A hash table is created on the resulting table and then a map-side join is executed with the *supplier* table. Then, a common join is executed between the table produced and the *lineitem* table. The common join is a MapReduce job that scans the two tables in the map phase, repartitions them over the shuffle phase on the join attribute, and finally performs the join in the reduce phase. This join runs for about 14880 seconds at the 16TB scale dataset. The map and shuffle phases run for approximately 12480 seconds. The output of this operation (TMP table) is then joined with the *orders* table using the common join mechanism. The running time of this join is 4140 seconds. Then, Hive performs a common join between the *customer* table and the output of the previous operation. During this operation the results are partially aggregated on the *n_name* attribute. This join runs for about 720 seconds. Finally, Hive launches two map-reduce jobs to perform the global aggregation as well as the order-by part of the query.

The reasons why Hive is slower than PDW when running Query 5 are described below.

First, the RCFile format is not a very efficient storage layout. We noticed that the read bandwidth when reading data from the RCFile is very low. For example, during the join between the *lineitem* table and the second temporary table that is created, the read bandwidth during the map phase is approximately 70 MB/sec and the map tasks were CPU-bound (the 8 disks used to hold the database can deliver, in aggregate, almost 800 MB/sec of I/O when accessed sequentially. Tests using the *testdfsio* benchmark showed that in our setup, HDFS delivers approximately 400 MB/sec of read sequential bandwidth).

Another important reason for PDW's improved performance over Hive is that PDW repartitions the intermediate tables so that the subsequent join operations in the query plan can be executed locally. This repartitioning step is generated because the PDW optimizer computes a query plan, and splits the query into sub-queries using cost-based methods that minimize network transfers. As a result, large base tables, like *lineitem*, are not shuffled. Hive on the other hand, does not use any cost-based model to optimize query execution. The order of the joins is determined by the way the user (in this case the Hive developers) wrote the query. This approach results in missing opportunities to optimize joins. For example, since the join order is determined by the way the query is written, the table produced by joining the *nation*, *region* and *supplier* table has to be joined with the *lineitem* table. The *lineitem* table is not "bucketed" on an attribute related to the *supplier* table. As a result, the join is executed using the expensive common join mechanism that repartitions both tables and joins them in the reduce phase. The running time of this task is higher than the total running time of the PDW query at the 16TB scale factor. Another, example is the join between the TMP table and the *orders* table. Notice that the TMP table is produced by a join operation on the *lineitem* table, which is bucketed on the *l_orderkey* attribute. However, the TMP table is not bucketed at all. As a result the join between the TMP table and the *orders* table cannot proceed as a bucketed map join and the common join mechanism is used.

## Query 19

Query 19 joins two tables (*lineitem*, *part*) and performs an aggregation on the output. This query contains a complex AND/OR selection predicate that involves both tables. This query was approximately 18 times faster in PDW at the 16TB scale factor. The plans produced by PDW and Hive are as follows:

**PDW:** PDW first replicates the *part* table at all the nodes of the cluster. This process is completed after 51 seconds. Then, it joins the *lineitem* table with the part table at each node, applies the selection predicate and performs a local aggregation operation; these three operations run for approximately 906 seconds. Finally, it performs a global aggregation of all the results produced by the previous stage.

**Hive:** Hive performs a common join between the *lineitem* and the *part* table. At the 16TB scale, this join operation runs for about 17540 seconds. The map and shuffle phases run for 14220 seconds. During the reduce phase, a partial aggregation is also performed. Then, Hive launches one more MapReduce job that performs the global aggregation. This job runs for about 25 seconds at the 16 TB scale factor.

As with query 5, PDW tries to avoid network transfers. For this reason, it replicates the small table (*part*), and then performs the join with the *lineitem* table locally. Hive, on the other hand, redistributes both the *part* and the *lineitem* tables and then performs the join in the reduce phase of the MapReduce job. Hive could have performed a map-side join instead of a common join, but it doesn't make that choice, probably because a hash table on the *part* table wouldn't fit in the memory assigned to each map task.

Similar arguments hold for other queries where PDW significantly outperforms Hive (e.g. Q7, Q8).

### 3.3.4.2 Scalability Analysis
As shown in Table 3, Hive scales well as the dataset size increases. In this section we analyze some queries where Hive scales sub-linearly when the dataset size increases by a factor of 4.

## Query 1

Query 1 scans the *lineitem* table and performs an aggregation followed by an order-by clause. The bulk of the time in this query is spent in the map phase of the MapReduce job that scans the *lineitem* table. The map tasks scan parts of the *lineitem* table and perform a map-side aggregation. Table 4 shows the total time spent in the map phase at each scale factor.

As shown in Table 4, when the dataset's size increases from 250 GB to 1TB, the map phase time increases by 2.3X. When the dataset increases from 1TB to 4TB the map phase time grows by a factor of 3.7. This factor becomes 4 when the dataset increases from 4TB to 16TB. The reason for this behavior is as follows:

The *lineitem* table contains 512 buckets based on the *l_orderkey* attribute, and it consists of 512 HDFS files (one file per bucket). We noticed that only 128 files out of the 512 contain data. The remaining 384 files are empty. According to the TPC-H specification, the *l_orderkey* attribute is sparsely populated (only the first 8 of every 32 keys are used). Hive uses hash partitioning to determine the bucket number that corresponds to each row. A hash function that assumes uniform distributions could have created this uneven distribution of data in buckets.

For the 250 GB dataset, 512 map tasks are launched (one per file). The map tasks that process non-empty files finish in approximately 75 seconds. The map tasks that process the empty files finish in 6 seconds. The total number of map tasks that can be run simultaneously on the cluster is 128 (128 map slots). Ideally, the first 128 map tasks would process non-empty files and complete in 75 seconds, and then 3 rounds would be needed to process the remaining empty files for a total time of 93 seconds. However, the total time is 148 seconds. This behavior happens because in the first round of map task allocation, both empty and non-empty files are processed. As a result, there is at least one map slot that processes two non-empty files, which then increases the total running time of the map phase.

For the larger dataset sizes, more than one map tasks process the non-empty files, and as a result, the ratio of the map tasks that do "useful" work over those that process empty files increases. For example, at the 1TB scale factor, 768 map tasks are launched (384 for the empty files and 384 for the non-empty files). As the datasets get bigger, the overhead introduced by the empty files reduces.

**Table 4. Total time for the map phase for Query 1**

| SF = 250 GB | SF = 1 TB | SF = 4 TB | SF = 16 TB |
|---|---|---|---|
| 148 secs | 339 secs | 1258 secs | 5220 secs |

Table 5. Time breakdown for Query 22

|  | SF = 250 GB | SF=1 TB | SF=4 TB | SF=16TB |
|---|---|---|---|---|
| **Sub-query 1** | 85 sec | 104 sec | 169 sec | 263 sec |
| **Sub-query 2** | 38 sec | 51 sec | 51 sec | 63 sec |
| **Sub-query 3** | 109 sec | 236 sec | 658 sec | 2234 sec |
| **Sub-query 4** | 654 sec | 735 sec | 797 sec | 813 sec |

## Query 22

Query 22 consists of four sub-queries in Hive. The average time spent in each sub-query for all scale factors is shown in Table 5.

Sub-query 1 scans the *customer* table, applies a selection predicate, and finally stores the output in a temporary table. The query consists of two MapReduce jobs at the first three scale factors and of one MapReduce job at the 16TB scale factor.

The first job executes the query (in a map-only phase), and outputs the result to a set of files (one per map task). The second job is a filesystem-related job that runs for 50 seconds at all the first three scale factors. This job stores the result of the previous query across fewer files. In the first MapReduce job, 200 map tasks are launched at the first three scale factors (one per *customer* bucket) and 600 map tasks when SF = 16TB. This is because at the 16TB each *customer* bucket consists of 3 HDFS blocks. The job time is 34 seconds when SF = 250 GB, 47 seconds when SF = 1TB, 102 seconds for the SF = 4TB and 263 sec when SF = 16TB. The job's running time does not increase by a factor of 4 as the dataset size increases by 4X. If we take a more careful look at the map phase of the job, we notice that each map task processes approximately 9.4 MB, 37 MB, 148 MB and 256 MB of compressed data at each scale factor. Each map task runs for about 9 seconds when SF = 250 GB, and 12 seconds when SF = 1000 GB. Since each map task processes a small amount of data (in the order of a few MB), the map task time does not scale linearly with the dataset size as the overhead associated with starting a new map task dominates the map task's running time.

Sub-query 2 consists of one MapReduce job that scans the output of the previous query, performs an aggregation, and stores the result into another table. When SF = 250 GB, two map tasks are launched and they finish after 12 seconds. When SF = 1TB, three map tasks are launched to process a total of 735 MB of data, and each task finishes after 27 seconds. When SF = 4TB, twelve map tasks are launched to process a total of 3 GB of data, and each one finishes in 27 seconds. Finally, at the 16TB scale factor 600 map tasks are launched to process a total of approximately 12 GB (each map task processes up to 102 MB and runs for at most 15 seconds). Observe that the running time of this query is the same at SF = 1000 and SF = 4000. The reasons for this behavior are:

1. The map task time is the same at both scale factors since each map task processes one HDFS block (256 MB).
2. Since the available number of map slots is 128, the map tasks launched at these two scale factors (3 and 12 map tasks respectively) can be executed in one round.

Sub-query 3 scans the *orders* table, performs an aggregation and stores the output in a temporary table. The *orders* table consists of 512 buckets on the *orderkey* attribute. Similar to the *lineitem* table, only 128 files actually contain data. The scaling behavior of this query is similar to that of Query 1 presented above.

Sub-query 4 performs two joins. The first one is executed between the outputs of Sub-query 1 and Sub-query 3. Hive attempts to perform a map-side join at all scale factors. However, the join always fails after about 400 seconds due to Java heap errors (this time varies slightly across all scale factors). Then, a backup task is launched that executes the join using Hive's common join mechanism. The join completes in 110 seconds, 150 seconds, 196 seconds and 300 seconds at the 4 scale factors. The reason for this scaling behavior is the small amount of data processed per map task (similarly to Sub-query 1). Similar observations hold for the remaining MapReduce jobs of this query (the second join, the group-by part, and the order-by part).

### 3.3.4.3 Discussion
Based on our analysis above, we now summarize the reasons that result in PDW outperforming Hive. These reasons are:

1. Although the RCFile format is an efficient storage format, it has a high CPU overhead. Previous work [16] has proposed a format that is more efficient than the RCFile storage layout and could potentially improve the performance of Hive.
2. Cost-based optimization in PDW results in better join ordering. In Hive, hand-written sub-queries and absence of cost-based optimization results in missed opportunities for better join processing.
3. Partitioning attributes in PDW are crucial inputs to the optimizer as it tries to produce plans with joins that can be done locally, and hence have low network transfer costs. In Hive although tables are divided into buckets, this information is not fully exploited by the optimizer. For example, the output of intermediate joins is not repartitioned (re-bucketed) so that the next join operator can be executed using a bucketed map join.
4. PDW replicates small tables to force local joins. Hive supports the notion of map-side join, which is similar to the replication mechanism of PDW. In a map-side join, a hash table is built on the smaller table at the Hive master node. The hash table is distributed to all the nodes using Hadoop's distributed cache mechanism. Then, all the map tasks load the hash table in-memory, scan the large table and perform a map-side only join. One disadvantage of this approach is that the hash table must fit in the memory assigned to each map task. As a result, there is a tradeoff between the number of map slots per node (which is determined by the user) and the memory available to each map task. This memory restriction is frequently the reason why this query plan fails. Another issue is that each new map task at a node has to load the hash table in memory from the local storage. (The hash table does not persist across map tasks on the same node). The authors in [18] present another alternative to map-side joins that avoids these issues.

    A related point is that although bucketing can improve performance by allowing bucketed map joins, the bucket should be small enough so that it can fit in the memory available to each map task. Having many small buckets at each table can help getting more bucketed map joins. However, when scanning tables that consist of many small buckets, the map task time can be dominated by the startup cost of the map task. Moreover, it's possible that the number of map tasks launched is high so multiple map rounds of short map tasks are needed to complete the scan (e.g. Sub-query 1 in Q22).
5. Finally, although Hive does not exploit bucketing as efficiently as partitioning is exploited by PDW, it is worth noting that unlike PDW, the buckets of two tables that are bucketed on the same attribute, are not guaranteed to be co-located on the filesystem (HDFS). Even if Hive is able to exploit the bucketing information more efficiently, absence of co-location would translate to network I/O, which in turn can significantly deteriorate performance [15].

Regarding scalability, Hive scales better than PDW (i.e. the scaling factors in the six right-most columns of Table 3 are lower for Hive) for the following reasons:

1. It has extra overheads for small datasets (e.g. overheads introduced by empty data files, startup cost of map tasks).
2. For some queries, increasing the dataset size does not affect the query time, since there is enough available parallelism to process the data (e.g. enough available map slots). Sub-query 2 of Q22 is such an example.
3. Some tasks take the same amount of time at all scale factors (filesystem-related job, map-side join fails after the same amount of time).

### 3.4 "Modern" OLTP Workload: MongoDB vs. SQL Server

In this section, we compare the performance of MongoDB and SQL Server in a cluster environment, using the YCSB data-serving benchmark [14].

**Table 6. YCSB benchmark workloads**

| Workload | Operations |
| --- | --- |
| A – Update heavy | Read: 50%, Update: 50% |
| B – Read heavy | Read: 95%, Update: 5% |
| C – Read only | Read: 100% |
| D – Read latest | Read: 95%, Append: 5% |
| E – Short ranges | Scan: 95%, Append: 5% |

#### 3.4.1 Workload Description

We used the YCSB benchmark, to evaluate our MongoDB implementation (Mongo-CS), the original MongoDB system (Mongo-AS), and our sharded SQL Server implementation (SQL-CS) on the "modern" OLTP workloads that represent the new class of cloud data-serving systems. The YCSB benchmark consists of five workloads that are summarized in Table 6. The YCSB paper [14] contains more details about the request distributions used by each workload. We have extended YCSB in the following two ways: First, we added support for multiple instances on many database servers, so as to measure the performance of client-sharded SQL Server (SQL-CS) and client-sharded MongoDB (Mongo-CS). Second, we added support for stored procedures in the YCSB JDBC driver.

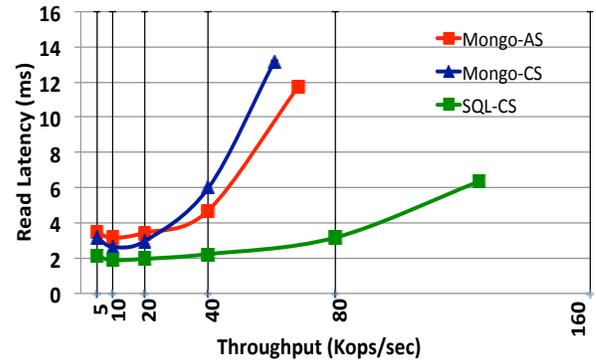

Figure 2. Workload C: 100% reads

We ran the YCSB benchmark on a database that consists of 640 million records (80M records per node). The dataset size per node is approximately 2.5 larger than the available main memory at each server machine. Each record in the database is 1024 bytes long and consists of one 24-byte key and 10 extra fields of 100 bytes each. All the fields as well as the key are stored as strings. Each key is generated by an integer, by using the string representation of the integer prefixing it with a sequence of '0', so that the total length of the key is 24 bytes. The data has an index on the record key, both in SQL Server and MongoDB. No other indexes were built in these systems. The record key is also used as the shard key for Mongo-AS.

We ran Mongo-AS and Mongo-CS in "safe" mode. This means that, after each write request, the client waits for a response from the server. This message shows that the server received the request and applied the write. However, there is no guarantee that the data was actually written to disk. We decided to not enable the "fsync" parameter that MongoDB provides, and as a result we do not wait for the writes to be flushed to disk before the response message is sent back to the client (this choice was made to improve the write performance in MongoDB).

While SQL Server supports ACID transaction semantics (at the default READ COMMITTED isolation level), the MongoDB experiments were run without durability support. The version of MongoDB that we used supports durability via write-ahead journaling. The journal is flushed to disk every 100 ms. This 100 ms delay means that the redo log by itself does not fully support durability, unless a commit acknowledgement is provided. For our experiments, we elected to run MongoDB without logging so that it doesn't pay any additional performance penalty.

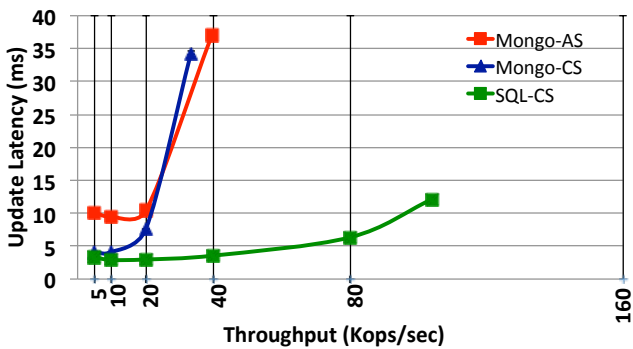

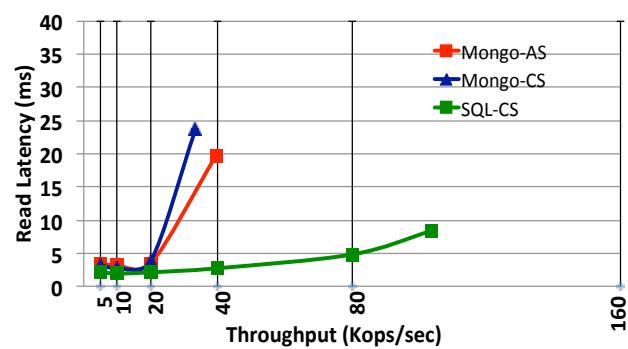

Figure 3. Workload B: 95% reads, 5% updates

In our setting, each client node runs 100 client threads for a total of 800 client threads (recall we have 8 machines dedicated to the clients). The five YCSB workloads are run sequentially, and before every run the main memory is flushed. After executing workloads D and E, which contain insertions and alter the record keys, the database is dropped and reloaded. Each read request reads all the record fields, and each update request updates only one field. Each scan request reads at most 1,000 records from the database. Finally, each append request inserts a new record in the database whose key has the next greater value than that of the last inserted key.

Each workload is run for 30 minutes. The values of latency and throughput reported are the average values over the last 10 minutes of execution, measured every 10 second interval. In the figures below, we also report the standard error across these 60 measurements.

### 3.4.2 Data Preparation
During the load phase, we used 8 client nodes, each running 16 client threads (as there are 16 hyper-threaded cores on each node). These threads are responsible for generating the correct keys and loading the data across the 8 server nodes.

Mongo-AS can automatically split and migrate data chunks across the shards by using a "balancer" process that takes care of load balancing. However, since the range and distribution of keys to be inserted are known in advance, we manually defined the boundaries for all of the initially empty chunks and spread them across the 128 shards of the cluster. Then, we started loading the data. In this way, the high cost of chunk migration across the shards is minimized. This technique is described in the MongoDB documentation [9]. This process resulted in an even distribution of the chunks across all the shards. The loading time with this strategy was 114 minutes.

The loading time for SQL-CS and Mongo-CS was 146 and 45 minutes respectively. The SQL-CS load time is higher than that of the Mongo-CS system because a bulk insert method was not used to load the data. Instead, every insertion was a separate transaction issued at the database.

### 3.4.3 Experimental Evaluation
The YCSB benchmark focuses on the latency of requests when the data-serving system is under load. However, as the load increases on a given system, the latency of requests typically increases since there is more contention for resources. In practice, the cloud service providers decide on an acceptable latency, and then provision enough servers to achieve the desired throughput. The YCSB benchmark aims to describe the tradeoffs between throughput and latency for each system by measuring latency as throughput is increased, until the point at which the system is saturated and throughput stops increasing. To run the benchmark, the (benchmark) user provides a target throughput as an input parameter, and the system returns the average latency as well as the actual throughput that is achieved. The user stops increasing the target throughput when the actual throughput that is achieved is lower than the target value.

Figure 2 shows the latency vs. throughput curve for the "Read-Only" workload (Workload C). The label on the x-axis shows the target throughput values provided by the user. Each data point corresponds to a pair of the actual throughput achieved and the average read latency for that throughput.

As shown in Figure 2, SQL-CS is able to achieve the highest throughput (125,457 ops/sec) with an average read latency of 6.4 ms. Mongo-AS and Mongo-CS were not able to reach the 80,000 ops/sec of target throughput and peaked at 68,533 and 60,907 ops/sec respectively. The average read latency values at the highest throughput achieved for Mongo-AS and Mongo-CS are 11.8 ms and 13.2 ms respectively. Moreover, SQL-CS has lower latency than the other systems for all the target throughputs. This workload is disk-bound is all the systems at the highest achievable throughput. However, the latency of each read request is higher with the Mongo-AS and the Mongo-CS systems compared to SQL-CS. We noticed, that SQL Server reads 8KB from disk for each request that leads to a buffer pool miss, whereas Mongo-AS and Mongo-CS read on average 32 KB from disk for each read request. Since the I/O activity pattern in this workload is largely random access, Mongo-AS and Mongo-CS waste disk bandwidth by reading in data that is not needed.

Figure 3 presents the latency vs. throughput curves for the "Read-Heavy" workload (Workload B). The workload consists of 95% reads and 5% updates. The left-hand curve presents latency results for the update operation as the target throughput increases, and the right-hand curve presents latency results for the read operation. The Mongo-DB systems cannot achieve the 40,000 ops/sec throughput target. Moreover, the update and read latencies increase abruptly (up to 24 ms. for read requests and 37 ms. for update requests) when the throughput increases from 20,000 to 40,000 ops/sec. However, SQL-CS is able to achieve 103,789 ops/sec with an update latency of 12 ms, and a read latency of 8.4 ms. This workload is disk-bound in each of these three systems. We noticed that each system achieves the same number of operations/sec as in Workload C. However, during checkpointing in SQL-CS, or when the MongoDB systems are flushing data to disk, the throughput decreases. For example when checkpointing was not happening, SQL-CS is able to reach on average about 15,000 ops/sec per server node (similar to workload C), but during the checkpointing interval the throughput decreases to 7,000-8,000 ops/sec. This is the reason why the maximum throughput achieved is lower in Workload B than with Workload C.

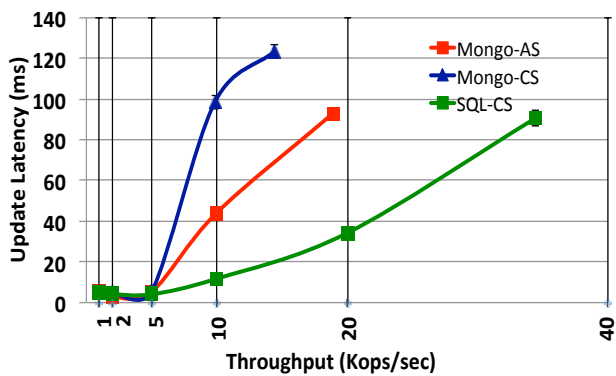 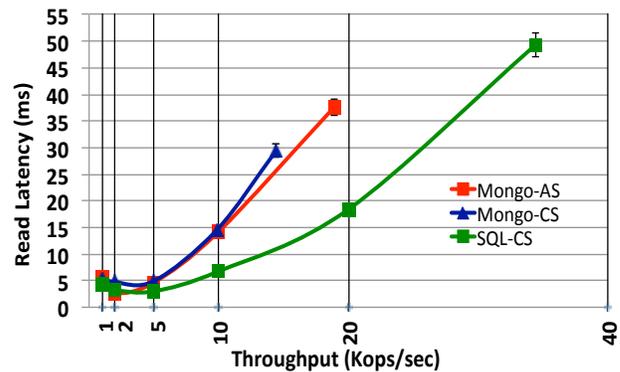

**Figure 4. Workload A: 50% reads, 50% updates**

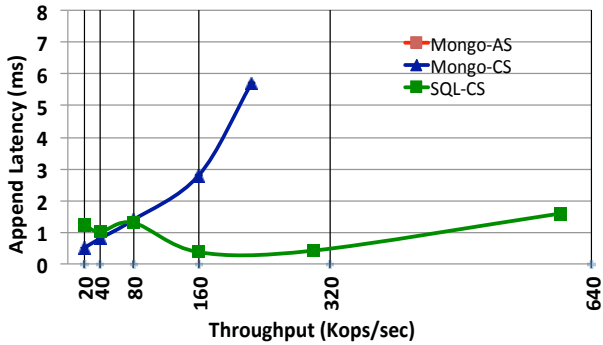
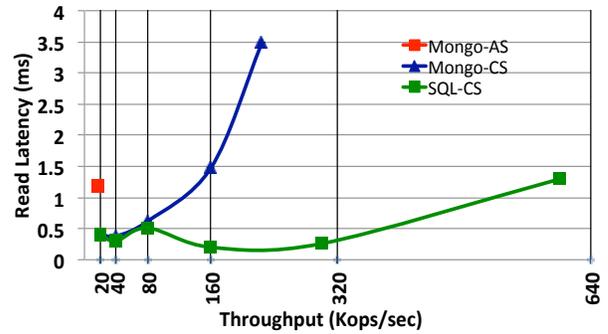

Figure 5. Workload D: 95% reads, 5% appends

Figure 4 describes the "Update-Heavy" workload (Workload A). This workload is similar to Workload B, but it contains a significantly larger fraction of updates (50% updates compared to 5% updates for Workload B).

Using the *mongostat* tool [10], we observed that the percentage of time that was spent at the global lock ranges from 25%-45% at each one of the 128 "mongod" instances. This percentage ranges from 4%-12% when running Workload B, which contains a smaller fraction of updates. Similarly, the increased locking activity in SQL-CS is the reason why both the read and update latencies are higher than those in Workload B. To verify this hypothesis, we reran the same workload but now using the "read uncommitted" isolation level and measured the read and update latencies. When the target throughput was 40,000 ops/sec, the average update latency was 69 ms. and the average read latency was 15 ms. The read latency is significantly lower now, compared to that of the previous experiment where the "read committed" isolation level was used. This can be attributed to the fact that the read operations are not blocked by the write operations and thus the waiting time is reduced.

Figure 5 shows the append latency and the read latency vs. throughput curves for Workload D. The read request distribution for Workload D is "Read Latest". This means that there is a high probability that a read request will read the latest item that was just inserted into the database.

We observed that in SQL-CS, 99.5% of the requests are to pages that are in the buffer pool. This means that the majority of the read requests do not hit the disk. Consequently, the read latencies for SQL-CS are in the order of a few microseconds. During the execution of this workload, SQL-CS is CPU-bound. SQL-CS has higher latencies for the low target throughput values (up to 80,000 ops/sec) compared to the greater throughput values (160,000 ops/sec and 320,000 ops/sec). This behavior happens because at the low throughput values, memory is not fully filled with useful data until after the 30-minute interval. For example, when the target throughput is 20,000 ops/sec, only 19.2 GB of the main memory is filled (of the 32 GB that is available) and as a result many read requests still incur a disk I/O.

Mongo-CS has high read and append latencies when it hits the highest achievable throughput (224,271 ops/sec) compared to SQL-CS. Interestingly, this workload is neither CPU-bound nor disk-bound (in MongoDB). Mongo-AS has a very high append latency (320 ms) for this workload when the target throughput is at 20,000 ops/sec, which is why this point does not appear in the graph in Figure 5. Moreover, Mongo-AS crashes when running this workload when the target throughput is set to a value greater than 20,000 ops/sec. After running the system with the debugger enabled, we observed that at some point the client machines wait for a response message from the server after an append request, but this message never arrives due to socket exceptions. For this reason, the clients stopped sending new requests to the servers and the throughput went down to 0 ops/sec.

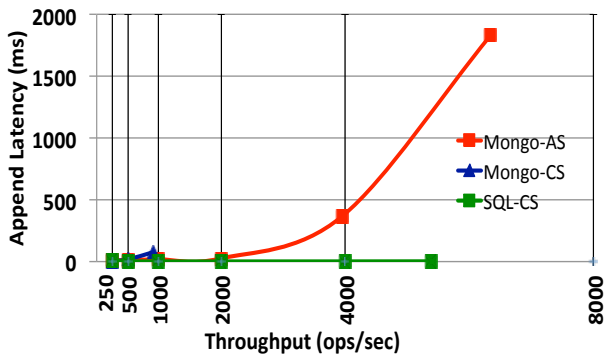
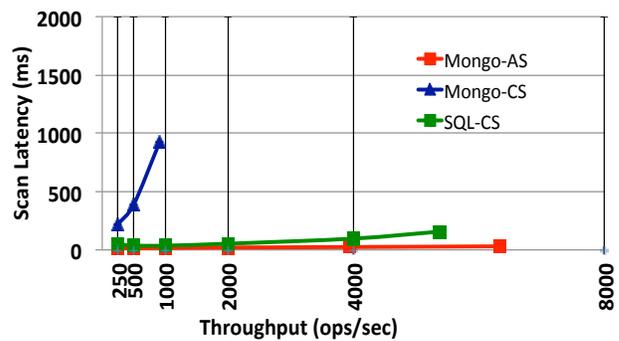

Figure 6. Workload E: 95% scans, 5% appends

Finally, Figure 6 shows the performance of the three systems on the "Short Ranges" workload (Workload E). All three systems are disk-bound when the servers hit their maximum throughput. As shown in the figure, Mongo-AS achieves the highest throughput (6,337 ops/sec) and has the lowest scan latency (30.4 ms). This behavior can be attributed to the fact that Mongo-AS uses range partitioning to distribute the data chunks across the servers, whereas both SQL-CS and Mongo-CS use hash partitioning. That means that Mongo-AS can determine, based on the range requested, which partitions contain the data and scan only those (typically one partition for each short range query) whereas SQL-CS and Mongo-CS need to scan as many partitions as needed until the appropriate records are found. However, Mongo-AS has a very high append latency (1832 ms) compared to SQL-CS (2 ms).

## 3.5 Discussion

In this section we compared a SQL system (PDW, SQL-CS) and a representative NoSQL system (Hive and MongoDB) on a DSS and an OLTP workload.

Our evaluation has shown that although NoSQL systems have significantly evolved over the past years, their performance still lags behind that of the relational database systems. On the one hand, the parallel database system (PDW) was approximately 9X faster than the MapReduce-based data warehouse (Hive) when running TPC-H at a 16TB scale, even when indexing was not used in PDW. The robust and mature cost-based optimization and sophisticated query evaluation techniques that are employed by the relational database system allow it to produce and run more efficient plans than the NoSQL system. The MapReduce-based systems could adopt these techniques to improve their performance.

Furthermore, SQL-CS was able to achieve higher throughput than the MongoDB for the same number of clients, and it had lower latency across for almost every single test of the YCSB benchmark. Interestingly, this is the case even when the NoSQL system did not provide any form of durability. This finding comes in contrast with the widely held belief that relational databases might be too heavy weight for this type of workload, where the requests consist of a single simple operation and do not require the complex transactional semantics that RDBMSs can handle.

## 4. CONCLUSIONS AND FUTURE WORK

Today there are a number of popular alternatives to using relational data processing systems, for both DSS workloads and the Web 2.0 data-serving workloads. While there are many complex factors that go into the choice of the system that gets deployed for specific data processing tasks (such as integration of the data processing system with an overall solutions stack, manageability, open-source vs. closed-source, etc.), one crucial aspect that is often a factor in choosing a data processing system is the performance of the system. In this paper, we examined this performance aspect of NoSQL and SQL systems using two benchmarks – the TPC-H benchmark and the YCSB benchmark. Our results find that the relational systems continue to provide a significant performance advantage over their NoSQL counterparts, but the NoSQL alternatives are competitive in some cases. The NoSQL and SQL systems also have different focuses on non-performance related features, such as data models (the NoSQL systems tend to have more flexible data models), support for auto-sharding and automatic load balancing and different consistency models. It is likely that in the future these systems will start to converge on the functionality aspects. An interesting direction for future work is to expand this work to other SQL and NoSQL systems and revisit the performance differences in a few years.

## 5. ACKNOWLEDGMENTS

This work was supported by a grant from the Microsoft Jim Gray Systems Lab to the University of Wisconsin-Madison. We would also like to thank Alan Halverson and Dimitris Tsirogiannis for their valuable input on this work.